\documentclass[a4paper,11pt]{article}
\usepackage{pos}
\usepackage{physics}
\usepackage{subcaption}
\usepackage{todonotes}
\usepackage{amsthm}
\usepackage{amsfonts}
\usepackage{bbold}
\usepackage{hyperref}
\usepackage{setspace}
\title{Towards precision charm physics with a mixed action}

\author*[a,c]{A. Conigli}
\author[c]{J. Frison}
\author[a,b]{G. Herdo\'iza}
\author[a,b]{C. Pena}
\author[a,b]{A. S\'aez}
\author[a,b]{J. Ugarrio}
\affiliation[a]{Instituto de F\'isica Te\'orica UAM-CSIC, c/ Nicol\'as Cabrera 13-15, Universidad Aut\'onoma de Madrid, E-28049 Madrid, Spain}

\affiliation[b]{Department of Theoretical Physics, Universidad Aut\'onoma de Madrid, E-28049 Madrid, Spain}

\affiliation[c]{John von Neumann Institute for Computing, DESY, Platanenallee 6, 15738 Zeuthen,
	Germany}

\emailAdd{alessandro.conigli@uam.es}

\abstract{We report on our first set of results for charm physics, using a mixed-action setup with maximally twisted valence fermions on CLS $N_f=2+1$ ensembles. This setup avoids the need of improvement coefficients to subtract $O(am_c)$ effects. The charm quark mass, $D$ and $D_s$ decay constants  are computed on a subset of CLS ensembles, which allows to take the continuum limit and extrapolate to the physical pion mass, and assess the scaling properties. Special attention is paid to the implementation of techniques to deal with systematic uncertainties. Our results show excellent prospects for high-precision computations on the full set of ensembles. }

\FullConference{%
	The 39th International Symposium on Lattice Field Theory,\\
	8th-13th August, 2022,\\
	Rheinische Friedrich-Wilhelms-Universität Bonn, Bonn, Germany
}

\begin{document}
	\maketitle
	
	\section{Introduction }
	\noindent
	The Standard Model of Particle Physics has proven to be one of the most promising and successful physical theories ever developed, yet it shows inconsistencies that might lead to new physics beyond. A promising landscapes to pursuit these searches is the hadronic flavour sector, where processes involving heavy quarks play a crucial role.  The investigation of these phenomena requires a non-perturbative framework such as Lattice Field Theory to perform those computations from first principles. 
	
	We introduced \cite{tmQCD} a mixed-action setup that targets the control of  systematics in heavy quark computations. Our approach employs CLS ensembles \cite{cls1,cls2} with open boundary conditions in the time directions, combined with a Wilson twisted mass action in the valence sector to compute $O(a)$-improved observables without improvement coefficients \cite{Bussone:2018ljj, Bussone:2019mlt}.  In \cite{Alejandro22} we review the scale-setting procedure of our setup, while in \cite{Gregorio22} we present preliminary results for the light quark masses. Here we focus on the study of charm physics observables,  and we discuss an update from \cite{Ugarrio:2018ghf, Conigli:2021} for the charm quark mass and for charmed mesons decay constants, presenting final results with the first generation of CLS ensembles at four values of the lattice spacing.
	
	\section{tmQCD mixed action}
	\label{sec:tmqcd}
	\noindent
	The set of gauge ensembles used in our study were produced within the CLS initiative \cite{cls1, cls2}.  
	The gauge action employed in CLS  ensembles  is the so-called tree-level improved L\"{u}scher-Weisz gauge action \cite{gauge_action}, while on the fermionic sector the action involves a Wilson Dirac operator for $N_f=2+1$ flavours \cite{fermion_action}, where the Sheikholeslami-Wolhert  $c_{\mathrm{sw}}$ coefficient is computed non-perturbatively \cite{csw}.
	In particular, this work is performed with the first generation of CLS $N_f=2+1$ ensemble, listed in Tab. \ref{tab:ens},  at four values of the lattice spacing ranging from 0.087 fm  down to 0.050 fm.
	
	The ensembles lie  along an approximate line of constant trace of the bare quark mass matrix,
	\begin{equation}
		\tr(M_q) = 2m_{q,u} + m_{q,s} = \text{const},
		\label{eq:tr_mass_matrix}
	\end{equation}
	where $m_{q,f} = m_{0,f} - m_{\text{cr}}$.
	As a result, changes in the sea quark masses keep constant the cutoff effects of $O(a\tr M_q)$ appearing in the Symanzik expansion of the bare coupling. 
	
	While approaching the physical point it is beneficial to consider a renormalised chiral trajectory in terms of the pion and kaon masses. Specifically, the following dimensionless quantities 
	\begin{equation}
		\phi_4 \equiv 8t_0 \bigg(
		\frac{1}{2}m_\pi^2 + m_K^2
		\bigg),  \qquad \phi_2 \equiv 8t_0 m_\pi^2,
		\label{eq:phi2_and_phi4_def}
	\end{equation}
	depending on the gradient flow observable $t_0$, are used to establish the chiral trajectory \cite{cls1}.
	In this setup, the value of $\phi_4$ has to be fixed to  its physical value$ \phi_4^{\mathrm{phys}} = 1.110(13)$ on each ensemble, where we have used the physical values of pion and kaon masses and the physical value of the gradient flow $t_0^{\mathrm{phys}}$ as determined in \cite{Bruno:2016plf}.  The light quark mass dependence of physical quantities is then encoded in the  $\phi_2$ hadronic combination and the physical point is reached via an extrapolation to $	\phi_2^\mathrm{phys} = 0.0796(19)$. In practice, deviations from $\phi_4^{\mathrm{phys}}$ have been observed to be non negligible. Therefore, in order to restore $\phi_4 = \phi_4^{\mathrm{phys}}$ on each ensemble, we apply small corrections  to the simulated bare quark masses. This mass shift is then applied to any observable by mean of a Taylor expansion as pioneered in  \cite{Bruno:2016plf}.
	
	The mixed action setup in the valence sector contains $N_f=2+1+1$ quark flavours regularised with a Wilson twisted mass Dirac operator at maximal twist to achieve automatic $O(a)$ improvement \cite{tmQCD}.  To achieve full twist and recover the unitarity of the theory we simultaneously  impose the bare standard mass matrix to be equal to the critical mass and we match the sea and valence by imposing the light pseudoscalar masses to be  equal in both sectors. More details on the full twist regime and the matching procedure can be found in \cite{Bussone:2019mlt,gregorio} and the more recent update \cite{Alejandro22}.

\begin{table}[h!]
	\begin{center}	
		\begin{tabular}{c  c  c  c  c  c  c}
			\hline
			\text{Id} & $\beta $ & $N_s$ & $N_t$ & $m_\pi$ [MeV] & $m_K$ [MeV] & $M_\pi L $\\
			\hline
			H101 & 3.4 & 32 & 96 & 420 & 420 & 5.8  \\
			H102 & 3.4  & 32 & 96 & 350 & 440 & 5.9 \\
			H105 & 3.4  & 32 & 96 & 280 & 460 & 3.9 \\
			\hline 
			H400 & 3.46 & 32 & 96 & 420 & 420 & 5.2  \\
			\hline
			N200 & 3.55 & 48 & 128 & 280 & 460 & 4.4  \\
			N202 & 3.55 & 48 & 128 & 420 & 420 & 6.5  \\
			N203 & 3.55 & 48 & 128 & 340 & 440 & 5.4  \\
			D200 & 3.55 & 64 & 128 & 200 & 480 & 4.2  \\
			\hline
			N300 & 3.70 & 48 & 128 & 420 & 420 & 5.1 \\
			J303 & 3.70 & 64 & 196 & 260 & 470 & 4.1  \\
			\hline
		\end{tabular}
		\caption{List of CLS $N_f=2+1$ ensembles used in the present study \cite{cls1}. $N_s$ and $N_t$ refer to the spatial and temporal extent of the lattice. Approximate values of the pion and kaon masses are provided. }
		\label{tab:ens}
	\end{center}
\end{table}

\subsection{Matching to the physical charm quark mass }
\label{sec:cqmatch}
\noindent
In our setup, the charm quark is not a dynamical fermion, hence its matching procedure requires a different strategy.  For this reason,  we simulate at three  values around the charm quark mass and the  matching can be performed by imposing that some charm observable $\phi_c^{(i)} = \sqrt{8t_0}M_H^{(i)}$   in terms of the reference scale $t_0$ matches to its physical value for each ensemble. We studied three  tuning strategies based on flavour-averaged, spin-flavour averaged combinations of meson masses and on the connected contributions of the $\eta_c$ mass:
\begin{eqnarray}
	M_H^{(1)} &=& \frac{1}{3} \big(2M_D + M_{D_s}\big), \nonumber
	\\
	M_H^{(2)} &=& M_{\eta_c},\nonumber
	\\
	M_H^{(2)} &=& \frac{1}{4} \big(M_H^{(1)} + 2M^*_{D}+ M^*_{D_s}\big).
	\label{eq:match}
\end{eqnarray}

However this method present some drawbacks, since it introduces the dependence on the physical scale $t_0^{\mathrm{phys}}$ at finite lattice spacing. Moreover, meson masses in each ensembles are computed at unphysical values of $\phi_2$ and they contains $O(a^2)$ cutoff effects. Therefore, we prefer to perform the charm quark matching jointly with the continuum-chiral extrapolation. We eventually parametrize the charm quark mass dependence of a given observable $\mathcal{O}_c(a, \phi_2, \phi_c^{(i)})$ and perform a combine fit to its physical value  $\mathcal{O}_c(0, \phi_2|_\mathrm{phys}, \phi_c^{(i)}|_{\mathrm{phys}})$.
In sections \ref{sec:cq_mass} and \ref{sec:fds_decays} we will describe the functional forms that we used for the charm quark mass and the $D_{(s)}$ meson decay constants respectively.

\section{Computational details of the observables }
\noindent
We have measured two-point correlation functions at zero-momentum to extract ground-state meson masses and decay constants  from the CLS $N_f=2+1$ ensembles listed in Tab. \ref{tab:ens}. 

The open boundary conditions in the Euclidean time direction modify the spectrum of the theory in the neighbourhood of the boundaries. To address the boundary effects, the sources of the two-point functions are set in the bulk, precisely in the middle of the lattice at  $y_0 = T/2$. In more detail, the two-point correlation functions are defined as
\begin{equation}
	f_{\mathcal{O}\mathcal{O}'}^{qs} = \frac{a^6}{L^3}
	\sum_{\vec{x}, \vec{y}}\big\langle
	\mathcal{O}^{qs}(x_0,\vec{x})
	\mathcal{O}^{'qs}(y_0,\vec{y})
	\big\rangle,
	\label{eq:corr_func}
\end{equation}
where $y_0$ and $x_0$ are the source and sink time coordinates respectively while $q$ and $s$ are flavour indices. The quark bilinear operators $\mathcal{O}^{rs}$ are defined in terms of the of the  Euclidean gamma matrices  combinations $\Gamma$ as 
\begin{equation}
	\mathcal{O}^{qs}(x) = \overline{\psi}^q(x) \Gamma \psi^s(x),
\end{equation}
where we use $\Gamma = \gamma_5$ for the pseudoscalar density $P$. The automatic $O(a)$-improvement of the twisted mass formulation allows us to extract meson decay constants from pseudoscalar matrix elements as discussed later in more detail.
In this work we rely on the generalized eigenvalue problem (GEVP) variational method to compute the  spectrum and  matrix elements of charmed mesons \cite{gevp,gevp1}. We refer to \cite{Conigli:2021} for further details on the extraction of the observables from the GEVP.

	\subsection{Charm quark mass}
	\label{sec:cq_mass}
	\noindent
	As metioned in the previous section, to achieve the maximal twist regime we have to set the standard bare quark mass to its critical value.  As a consequence,
	PCAC quark masses entering the Ward identites vanish and all the physical information is encoded in the twisted mass matrix $\boldsymbol{\mu}_0 = \text{diag}(+\mu_{0,l},- \mu_{0,l},- \mu_{0,s},+ \mu_{0,c})$.
	Therefore we define the renormalised  charm quark mass in the twisted mass mixed action formulation as
	\begin{equation}
		\mu_c^{R} = Z_P^{-1}(g_0^2, a\mu_{had}) \mu_c(1+a\bar{b}_\mu \tr M_q) + O(a^2),
		\label{eq:renorm_charm_quark}
	\end{equation}
	where $\bar{b}_\mu = O(g_0^4)$ start at two-loop in perturbation theory \cite{Bussone:2018ljj}.
	At full twist, the definition in Eq. \ref{eq:renorm_charm_quark} is free from cutoff effects proportional to $O(a\mu)$ coming from the valence, while cutoff effects coming from the  sea light quark matrix can contribute but are highly suppressed by the fourth power of the coupling. In what follows we neglect these $O(a\tr M_q)$ effects  and  our assumption is supported by continuum limit extrapolation, where we  observe behaviour consistent with $O(a)$ improved quantities.
	
	The charm quark mass estimator in Eq. \ref{eq:renorm_charm_quark} is then combined with the  flavour independent running factor 
	\begin{equation}
		\frac{M}{\overline{m}(\mu_{\mathrm{had}})} = 0.9148(88)
	\end{equation}
	computed in \cite{ren_const} using  the Schr\"{o}dinger Functional scheme for $N_f=3$ massless flavours  to defined the Renormalisation Group Invariant (RGI)  charm quark mass 
	\begin{equation}
		M_c^{\mathrm{RGI}} =   \frac{M}{\overline{m}(\mu_{\mathrm{had}})}Z_P^{-1}(g_0^2, a\mu_{had}) \mu_c .
		\label{eq:rgi_quark_mass}
	\end{equation}

	\subsection{$D_{(s)}$ meson decays}
	\label{sec:fds_decays}
	\noindent
	Together with the charm quark mass, in  this work we focus on a precise determination of the $D$ and $D_{s}$ meson decay constants, relevant for electroweak leptonic decays.
	The matrix element that mediate the weak interaction transition defines the decay constant $f_{qr}$ as
	\begin{equation}
		\abs{\bra{0} \mathcal{A}_0^{qr} \ket{P^{qr}(\mathbf{p}=0)} } = \frac{f_{qr}M_{qr}}{\sqrt{2M_{qr}L^3}},
	\end{equation}
	where the factor  $1 / \sqrt{2M_{qr}L^3}$ is the  relativistic normalization of a single particle state, while $\ket{P_{qr}}$ denotes the ground state for a pseudoscalar meson.
	
	At full twist the symmetries of the Wilson twisted mass formulation relate the physical axial current  to the vector current in the twisted basis for non-diagonal flavours that mixes up-type and down-type quarks,
	\begin{equation}
		\mathcal{V}_\mu^{qr} = - i \mathcal{A}_\mu^{qr}, \quad \mu_q > 0 >\mu_r.
	\end{equation}
	Moreover, the conservation of the  Ward identity on the lattice 
	\begin{equation}
		\bra{0} \partial_0^*V_0^{qr}\ket{P^{qr}(\mathbf{p}=0)} = i(\mu_q - \mu_r) \bra{0} P^{qr}\ket{P^{qr}(\mathbf{p}=0)},
	\end{equation}
	implies that the point-split current $\tilde{V}_\mu^{qr}$ renormalizes as in the continuum with a trivial factor $Z_{\tilde{V}}=1$. As a result, the renormalization constants $Z_\mu$ and $Z_P$ for the twisted mass and  pseudoscalar density respectively fulfil the condition
	\begin{equation}
		Z_\mu  = Z_P^{-1}.
	\end{equation}
	Therefore  the pseudoscalar decay constants in the Wtm formulation at full twist renormalize trivially, and  we are able to compute decay constants just with pseudoscalar-pseudoscalar correlators.  Eventually, we define the $f_D$ and $f_{D_{s}}$ decay constants as 
	\begin{equation}
		f_{D_{(s)}} = \sqrt{\frac{2}{L^3 m_{D_{(s)}}^3 }} (\mu_c + \mu_{l(s)}) 
		\bra{0} P^{c, l(s)}\ket{D_{(s)}},
		\label{eq:decay_const}    
	\end{equation}
	were we extract the relevant matrix elements from the eigenvectors of the GEVP at large distances where the exponential decay is dominated by the ground state, as explained in \cite{Conigli:2021}.
	
	\subsection{Chiral-continuum extrapolations}
	\noindent
	Once we have determined the ground state meson masses and decay constants from the GEVP, we eventually go ahead with a combined chiral-continuum extrapolation. 
	The  approach to the   physical  light quark mass dependence is controlled by the hadronic quantity  $\phi_2 = 8 t_0 m_\pi^2$ only, as we already fixed the  strange quark by taking $\tr M_q=\mathrm{const}$ (see Sec. \ref{sec:tmqcd}), while the interpolation to the physical charm quark mass is monitored by $\phi_c^{(i)} = \sqrt{8t_0} m_H^{(i)}$. As discussed in Sec. \ref{sec:cqmatch}, we studied three different matching conditions based on the meson masses combinations reported in Eq. \ref{eq:match}.
	
	We perform the extrapolations for the   renormalised charm quark mass of Eq. \ref{eq:renorm_charm_quark}  and the charm-light meson decay constants defined in Eq. \ref{eq:decay_const}. All the observables are made dimensionless through the factor  $\sqrt{8t_0}$ and the physical units are restored after the extrapolation to the physical point by dividing for  $\sqrt{8t_0^{\mathrm{phys}}}$.
	In the analysis we include all the ensembles listed in Tab. \ref{tab:ens}.
	
	For the lattice spacing dependence of the observables  we assume the leading cutoff effects to be of $O(a^2)$ as the mixed action at full twist ensures the absence of $O(a)$ cutoff effects\footnote{Apart from residual lattice artifacts proportional to the sea light quark masses. As explained in \cite{Bussone:2018ljj} these effects are negligible at the current precision. } and the relevant $O(a)$ improved renormalization constant are know non-perturbatively from \cite{ren_const}. Eventually our general ansatz for the lattice spacing dependence is parametrised by
	\begin{equation}\label{eq:cutoff_dependence}
		c_\mathcal{O}(\phi_2, \phi_H, a) = \frac{a^2}{8t_0} \big(
		c_1 + c_2\phi_2 + c_3\phi_H^2
		\big)
		+
		\frac{a^4}{(8t_0)^2}\big(
		c_4 + c_5\phi_H^4
		\big).
	\end{equation}
	Here we  allow for cutoff terms describing the higher  $O(a^4)$ effects and we also consider cutoff effects proportional to the light quark masses. In the twisted mass formulation of LQCD at maximal twist, all the odd powers of the lattice spacing are suppressed.

	The continuum behaviour is governed by the quantity $\sqrt{8t_0}\mathcal{O}^{\mathrm{cont}}$, whose definition  is observable-dependent. More details on this quantity for the charm quark mass and the $f_{D_{s}}$ decay constants  are given in the following.
	Finally to arrive at a combined model we follow a similar strategy as proposed in \cite{charm_quark} by either adding linearly or multiplying non-linearly the continuum and the lattice spacing dependencies
	\begin{eqnarray}\label{eq:tot_model}
		\sqrt{8t_0}\mathcal{O}^{\text{linear}}(\phi_2,\phi_H, a) &=&
		\sqrt{8t_0}\mathcal{O}^{\text{cont}}(\phi_2,\phi_H,0) + c_\mathcal{O}(\phi_2,\phi_H,a),
		\\
		\sqrt{8t_0}\mathcal{O}^{\text{non-lin}}(\phi_2,\phi_H, a) &=& 
		\sqrt{8t_0}\mathcal{O}^{\text{cont}}(\phi_2,\phi_H,0) \big(1+ c_\mathcal{O}(\phi_2,\phi_H,a)\big). \nonumber
	\end{eqnarray}

	In order to estimate the systematic effects arising from the model selection, we study all the possible combinations of the coefficients $c_i$ in Eq. (\ref{eq:cutoff_dependence}), for a total of 64 different models for each matching condition.
	To estimate the fit  parameters of all the possible models we use a $\chi^2$ minimization scheme.   Since we are dealing with highly correlated data the uncorrelated $\chi^2$ does not yield reliable estimates of the fit paramaters. In practice we bypass this problem by employing the so-called $\chi^2$ expected, $\chi^2_{\text{exp}}$, as estimator for the goodness of a fit \cite{chiexp}. Eventually, to extract a final result within each category, we employ a weighted model average by mean of an Information Criteria (IC)  \textit{à la} Akaike as proposed in \cite{Jay:2020, charm_quark}. 
	 We therefore introduce the IC coefficient for each functional form as  \cite{charm_quark}
	\begin{equation}
		\mathrm{IC} = \frac{\chi^2}{\chi^2_{\mathrm{exp}}}(N-k) + 2k + \frac{2k^2+2k}{N-k-1},
	\end{equation}
	where N is the number of data points  and k the number of parameters in the model. More details on the model selection procedure we employ are given in \cite{Conigli:2021}, while in \cite{Julien22} we study continuum scaling with a detailed comparison between  Wilson and twisted mass valence quarks.

	\section{Results}
	
	\subsection{Charm quark mass}
	\noindent
	We model the  continuum  dependence of the dimensionless  renormalised charm  quark mass with  the functional form 
	\begin{equation}\label{eq:mc_cont_dep}
		\sqrt{8t_0}M_c^{\text{cont}}(\phi_2, \phi_H,0) = p_0 + p_1\phi_2 + p_2\phi_H,
	\end{equation}
	where the chiral extrapolation and the matching to the physical charm are governed again by  $\phi_2$ and $\phi_H$ respectively. 	Cutoff effects are described by Eq. \ref{eq:cutoff_dependence} and we eventually arrive at a combine model as in Eq. \ref{eq:tot_model}.
	We recall that our data are classified in three categories, according to whether the charm is fixed to its physical value via the flavour-averaged, the spin flavour-averaged combinations or the $\eta_c$ mass. 
	We stress out that the light quark mass dependence is dominated by the sea pion mass $\phi_2$ only, since the mass shift corrections to the chiral trajectory ensure the kaon masses to be fixed by the condition $\tr(M_q^R)=\text{const}$.
	
	We observe that results coming from the flavour-averaged and the $\eta_c$ matching conditions are compatible, while the spin flavour-averaged combination is not well under control. In particular it  exhibits a higher value of the $\chi^2/\chi^2_{\mathrm{exp}}$, which is presumably a reflection of a poor control of the vector states.  Therefore we chose to exclude it from the weighted average, although  it would be automatically suppressed by the information criteria.

	\begin{figure}
		\begin{subfigure}{\textwidth}
			\centering
			\includegraphics[scale=0.28]{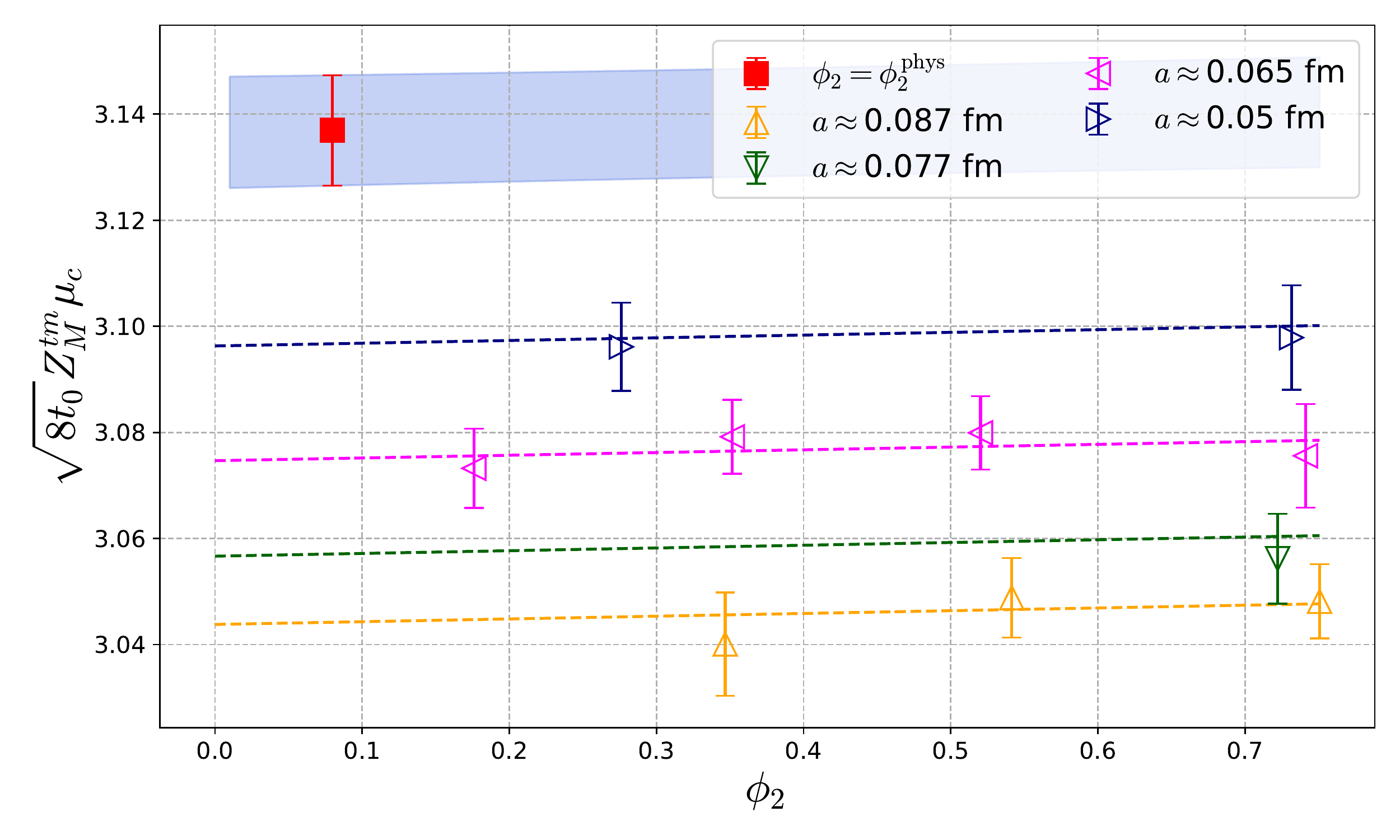}
			\includegraphics[scale=0.31]{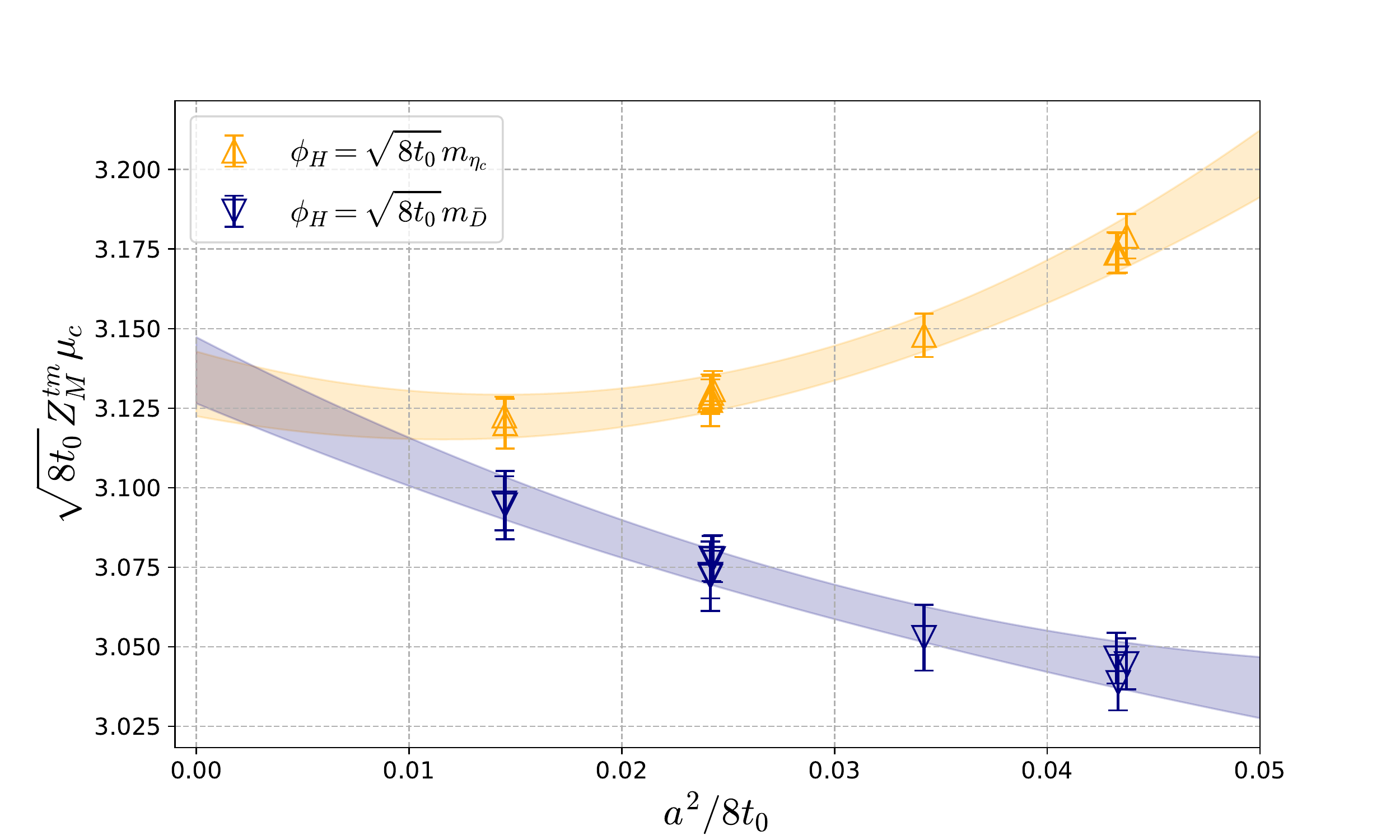}
		\end{subfigure}
	\caption{Comparison of some of the best fits according to the IC. \textit{Left:} chiral approach to the physical point of the charm quark mass for the flavour-averaged matching condition. The dashed lines corresponds to the chiral trajectories at finite lattice spacing, while the blue shaded band is a projection to the continuum limit fit. The red point represent the fit result at the physical point. \textit{Right:}  continuum limit behaviour of the charm quark mass for the flavour-averaged (blue) and the $\eta_c$ (yellow) matching conditions.  Data points are projected to the physical pion mass and the physical charm quark mass. 
	}
	\label{fig:mc_fit}
	\end{figure}
	In Fig. \ref{fig:mc_fit} we show  some of the best fits  for the chiral-continuum  extrapolation according to the IC for the renormalised charm quark mass in unit of $\sqrt{8t_0}$.  We notice that $O(a^4)$ cutoff effects have to be taken into account in both the flavour-averaged and  the $\eta_c$ matching prescriptions for a good description of the data, especially for coarsest values of the lattice spacing.

\begin{figure}
		\centering
		\begin{subfigure}{\textwidth}
			\centering
			\includegraphics[scale=0.32]{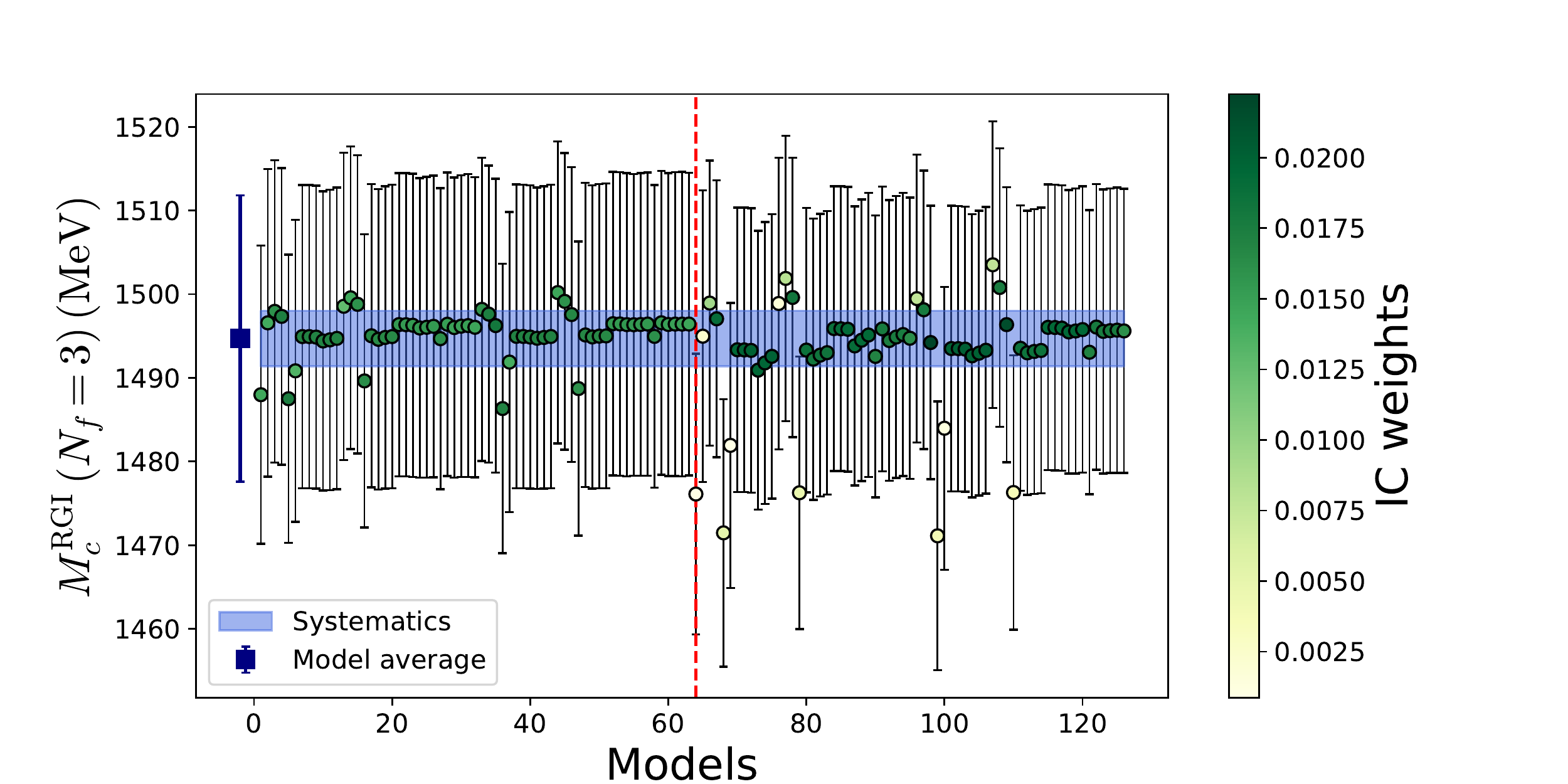}
			\includegraphics[scale=0.33]{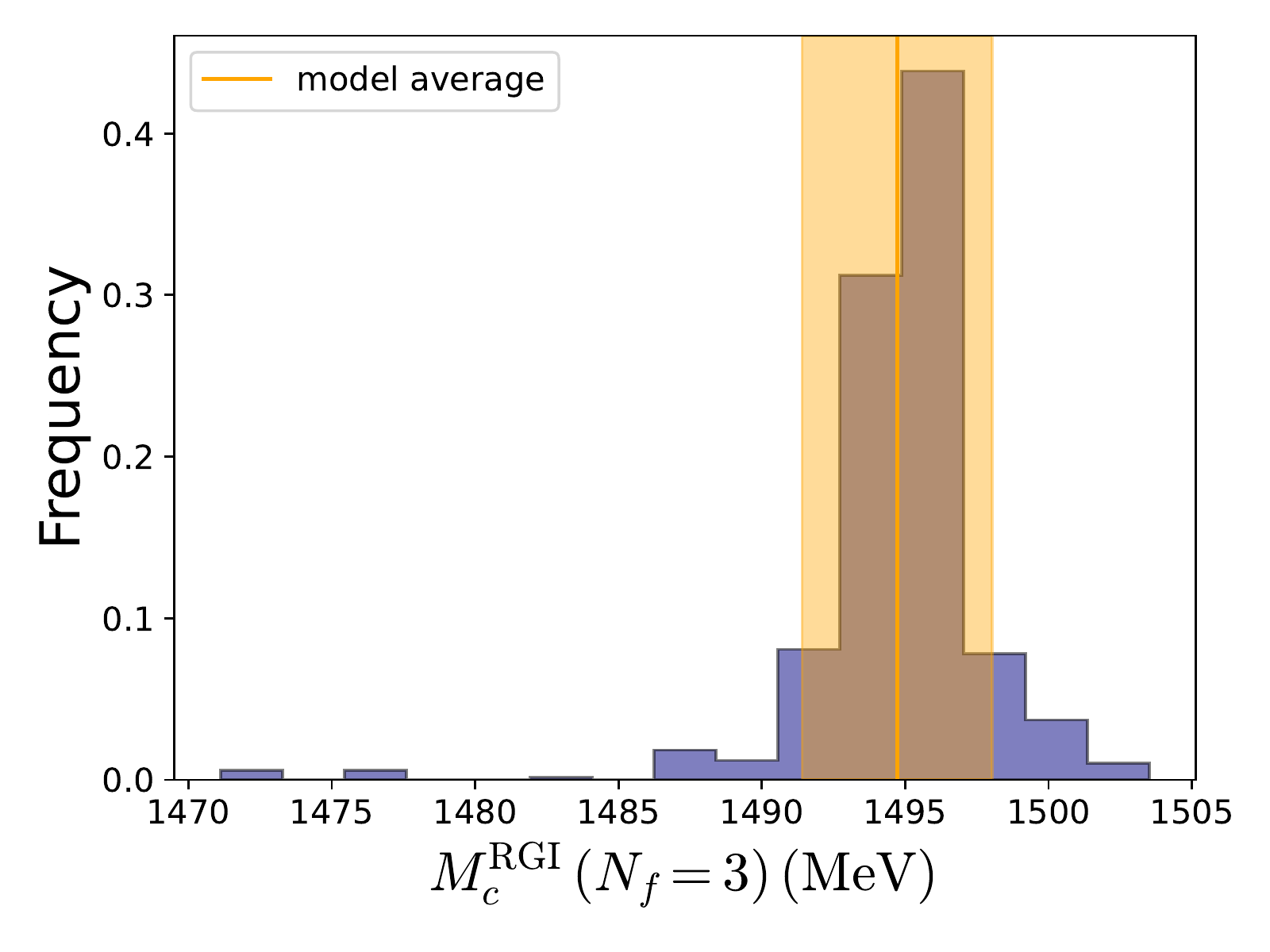}
		\end{subfigure}
	
	\caption{ Model average procedure for  $M_c^{\mathrm{RGI}}(N_f=3)$. \textit{Left}: summary of all the results coming from different fit models. Results on the left side of the red vertical line correspond to the flavour-averaged matching conditions, while the one on the right refers to the $\eta_c$ matching. The opacity of each point is associated to its weight on the model average. The blue shaded band represent the systematic error and the blue point the final result coming from the model average.  \textit{Right:}  weighted histogram collecting all results from the two different matching procedure. The yellow band represents the systematic error arising from the model selection. }
\label{fig:model_av_procedure}
	\end{figure}

	In Fig. \ref{fig:model_av_procedure} we summarise the model average procedure showing all the fit results coming from different models. We observe that they are mildly scattered around the model averaged value. Therefore our charm quark mass determination is dominated by the statistical error. Eventually, to extract the RGI mass we employ Eq. \ref{eq:rgi_quark_mass}. Following the analysis procedure described above, we quote as a final result for the RGI charm quark mass
	\begin{equation}
		M_c^{\mathrm{RGI}} = 1494(17)(3) \ \mathrm{MeV},
	\end{equation}
	where the first error is statistical and second accounts for the systematic arising from de model selection.  The dominant contribution to the error comes from the non-perturbative renormalization group running factor of Eq. \ref{eq:rgi_quark_mass} that connects renormalised quark masses to their RGI counterpart. Moreover, the second most dominant contribution comes from the scale setting procedure and the physical value of $t_0^{\mathrm{phys}}$. This tells us that the major error sources actually come from external inputs, while the statistical uncertainty arising from the correlation  functions is by far subleading.  In the first pie-chart in  Fig. \ref{fig:piechart} we report the error budget in our  $M_c^{\mathrm{RGI}}$ computation.

	\subsection{$D_{(s)}$ meson decays}
	\noindent
	The decay constants $f_D$ and $f_{D_s}$ are computed similarly to the charm quark mass. The continuum behaviour of the dimensionless renormalized decay constants can be inferred from Chiral Perturbation Theory with heavy quarks \cite{hqet1,hqet2}.   Therefore  we perform global fits for both $f_D$ and $f_{D_s}$ including chiral logarithm corrections
	\begin{eqnarray}
		f_D &=& p_0 + p_1\phi_2 + \frac{p_2}{\sqrt{\phi_H}}
		+p_3\bigg(
		3\mu_\pi + 2 \mu_K + \frac{1}{3}\mu_\eta  
		\bigg),
		\nonumber
		\\
		f_{D_s} &=& p_0 + 2p_1(\phi_4-\phi_2) + \frac{p_2}{\sqrt{\phi_H}}
		+p_3\bigg(
		4 \mu_K + \frac{4}{3}\mu_\eta\bigg),
	\end{eqnarray}
	where
	\begin{eqnarray}
		\mu_\pi &=& \phi_2 \log(\phi_2), \nonumber
		\\
		\mu_K &=& \bigg(\phi_4 - \frac{1}{2}\phi_2\bigg)\log(\phi_4 - \frac{1}{2}\phi_2), \nonumber
		\\
		\mu_\eta &= &\bigg(\frac{4}{3}\phi_4 - \phi_2\bigg)\log(\frac{4}{3}\phi_4 - \phi_2). \nonumber
	\end{eqnarray}
	Here $\phi_2$ and $\phi_4$ are the usual hadronic combinations defined in Eq. \ref{eq:phi2_and_phi4_def}.  For cutoff effects we consider a similar parameterization as in Eq.  \ref{eq:cutoff_dependence}. However, we observe that  non-linear terms describing cutoff effects turn out to be highly unstable. Hence, we only consider the linear combination for the decay constants, ending up with a total of 32 different models for each matching condition.	
	\begin{figure}
		\begin{subfigure}{\textwidth}
			\centering
			\includegraphics[scale=0.28]{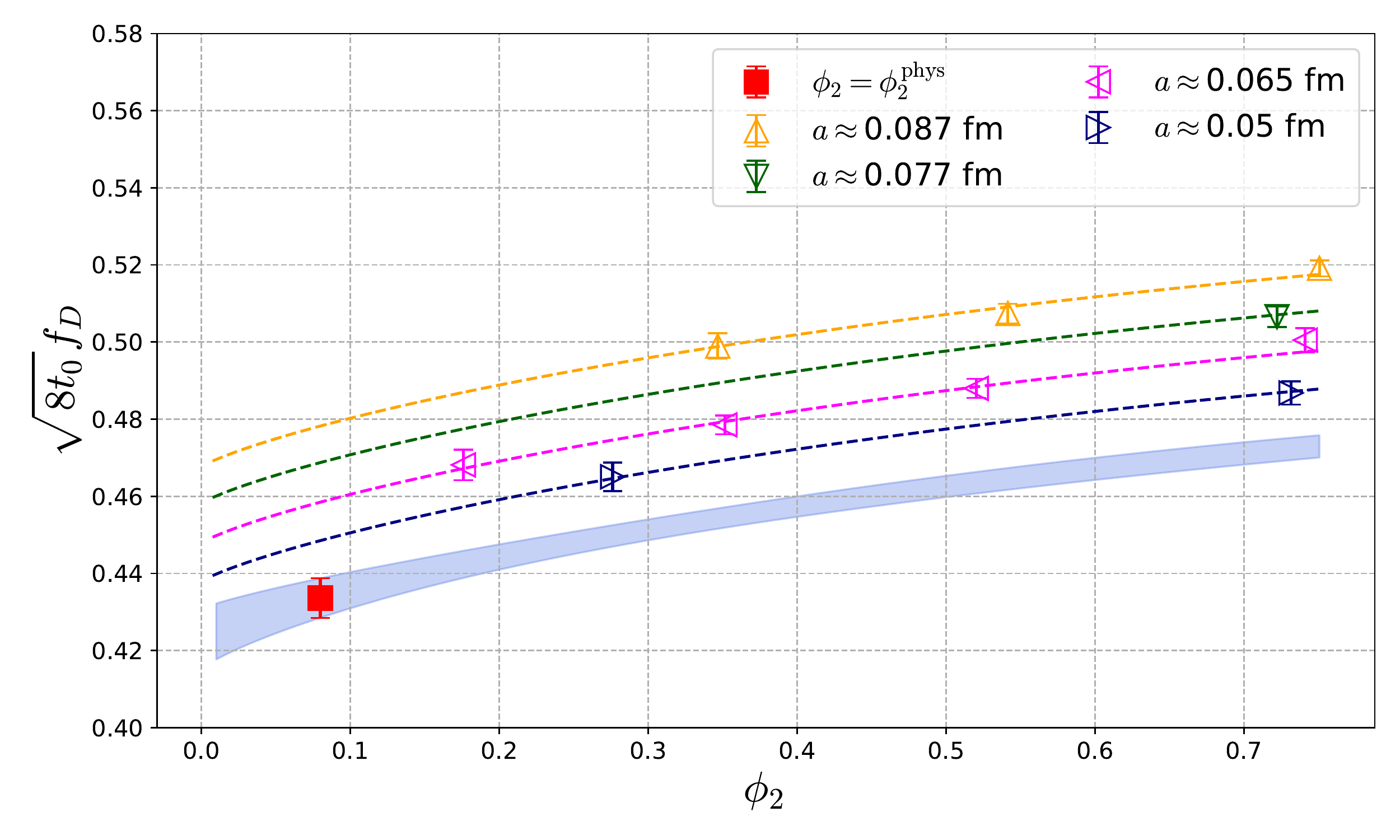}
			\includegraphics[scale=0.28]{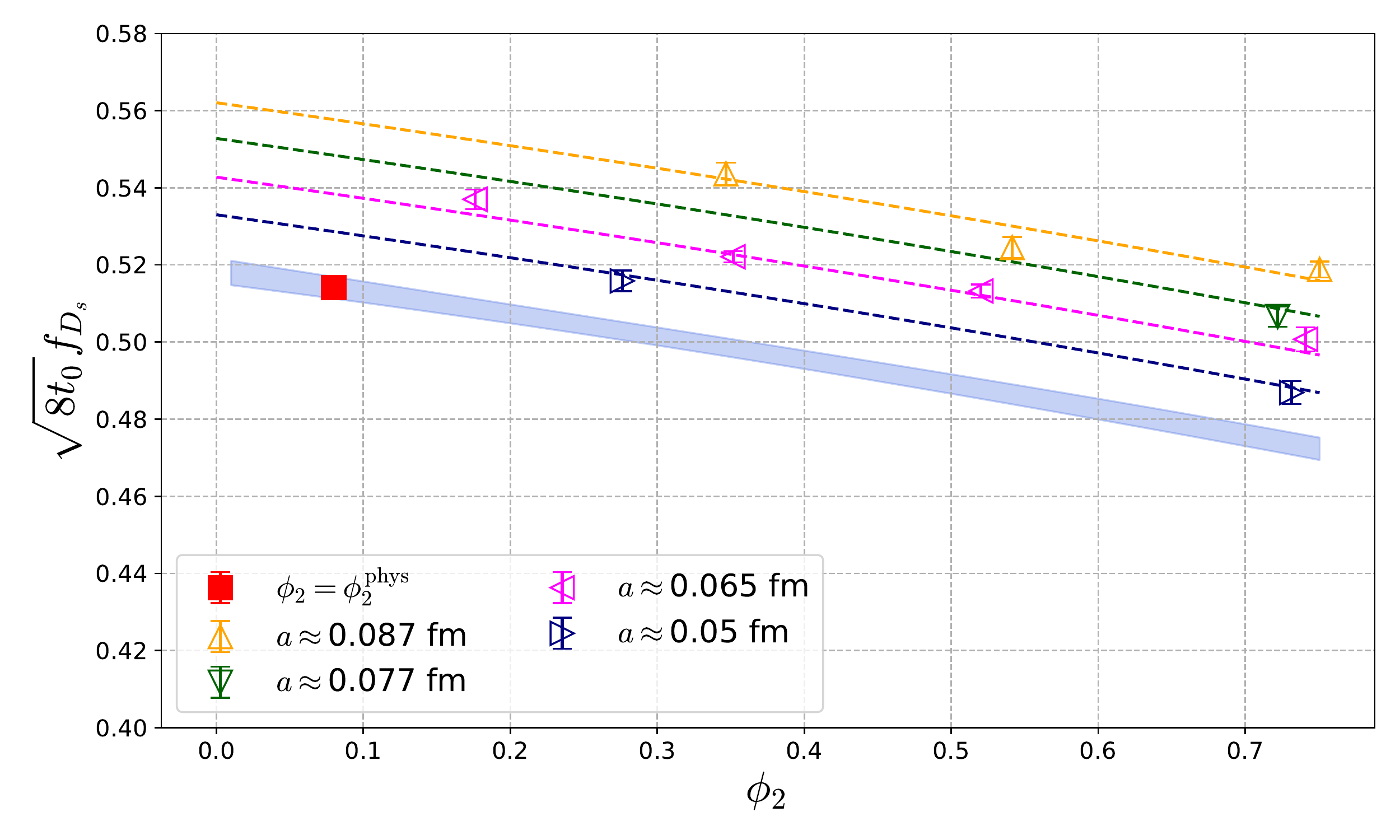}	
	\end{subfigure}
	\caption{Chiral behaviour  of some of the best fits according to the IC. \textit{Left:} $f_D$ decay constant. \textit{Right:}$f_{D_s}$ decay constant. Here the dashed lines show the chiral trajectories at finite lattice spacing, while the blue shaded bands are the projections to the continuum limit. Finally, the red points denote results at the physical point.}
	\label{fig:fit_fds}	
\end{figure}

\begin{figure}
		\centering
		\includegraphics[scale=0.39]{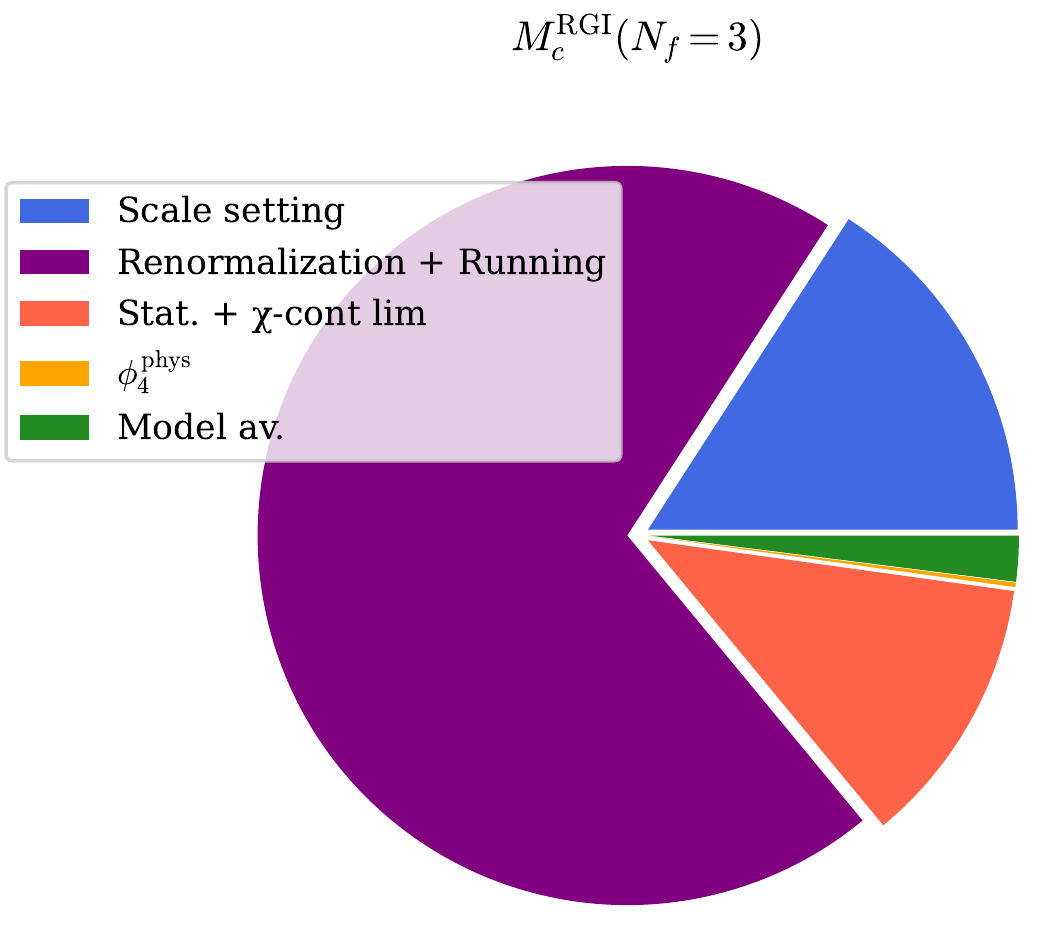}
		\hspace{0.4cm}
		\includegraphics[scale=0.39]{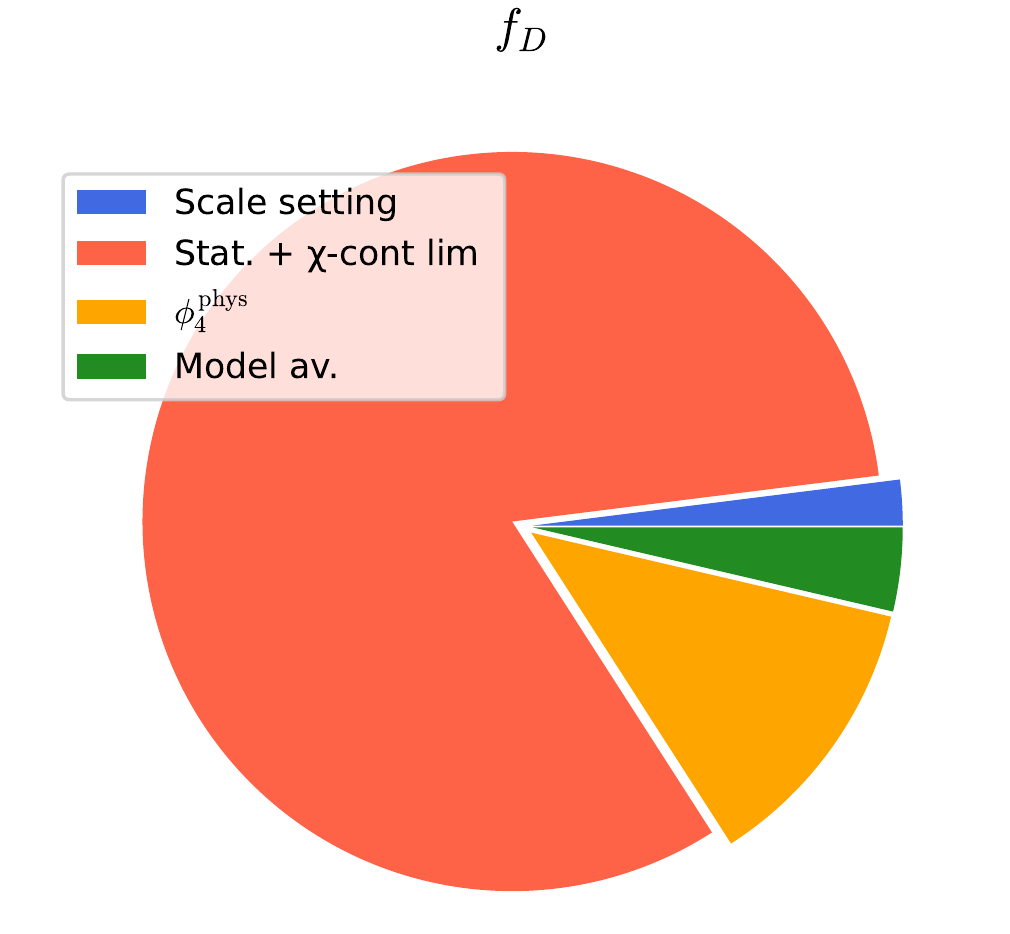}
		\includegraphics[scale=0.39]{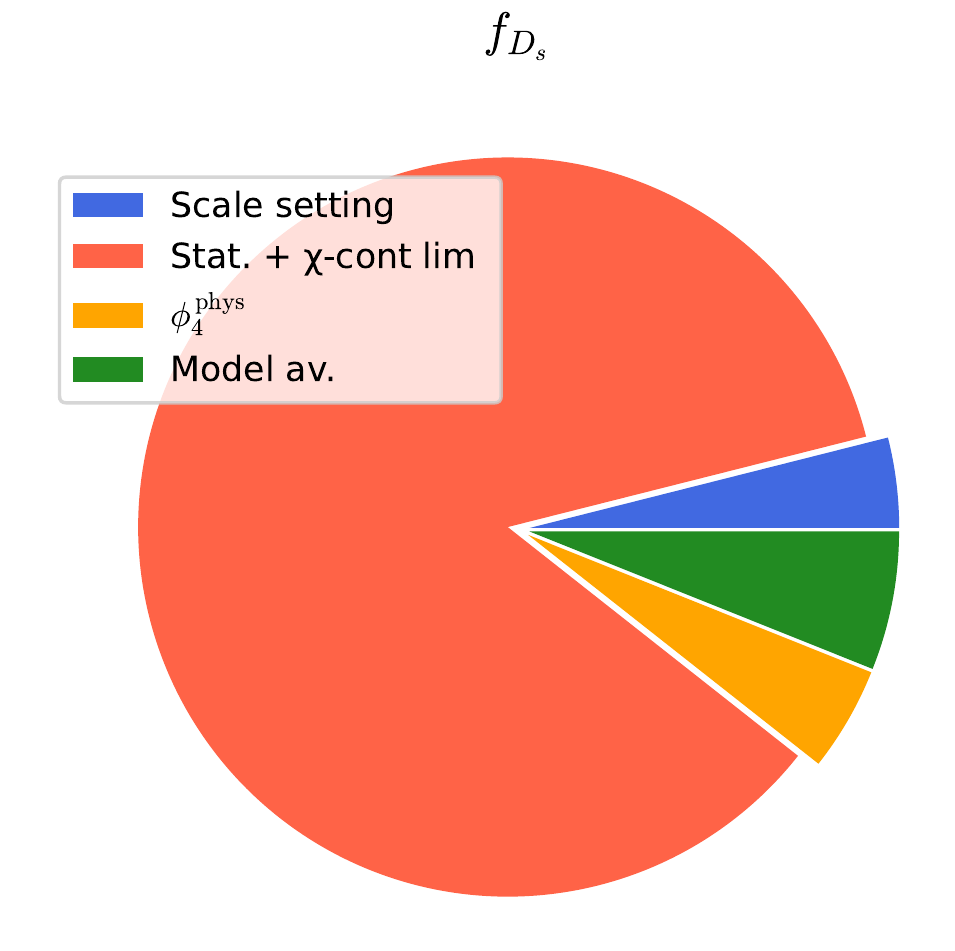}
		\caption{Dominant error contributions to the final results for $M_c^{\mathrm{RGI}}$ (\textit{left}), $f_D$ (\textit{center}) and $f_{D_s}$ (\textit{right}).  }
		\label{fig:piechart}
	\end{figure}

In Fig. \ref{fig:fit_fds}  we report the best fits for $f_D$ and $f_{D_s}$ respectively as given by the IC for different matching categories. We observe that in the case of the decay constants the spin-flavoured average matching condition turns out to be stable with reasonable values of the $\chi^2$ such that results coming from  this category are not suppressed by the model average procedure. Therefore, we include this matching condition in our average. Eventually from the  weighted  average for different models  we quote as preliminary results for the decay constants
\begin{eqnarray}
	f_D &=& 208.0(5.7)(1.9) \ \mathrm{MeV}, \nonumber
	\\
	f_{D_s} &=& 244.7(5.2)(0.9) \ \mathrm{MeV}, \nonumber
	\\
	f_{D_s} / f_D &=& 1.1709(83).
\end{eqnarray}
where for $f_{D_{(s)}}$ the first error is statistical and the second is the systematic, while for the ratio we only quote the statistical error. In Fig. 	\ref{fig:piechart} we plot the different error contributions to the $D_{(s)}$ meson decay constants.  In contrast to $M_c^{\mathrm{RGI}}$, the most dominant error source come from the statistical error of the correlation functions for both $f_D$ and $f_{D_s}$,  with the second subleading contribution being the external input $\phi_4^{\mathrm{phys}}$. 
	
	\section{Conclusions}
	\noindent
	We have presented an update on \cite{Conigli:2021} showing our most recent results for the RGI charm quark mass and the $D_{(s)}$ meson decay constants from a tmQCD mixed-action setup at full twist using a subset of CLS $N_f=2+1$ ensembles. Since our charm quark mass  error  budget is dominated by external inputs, there is no room for significant improvement in the precision. On the other hand, $f_{D_{(s)}}$ decay constants are dominated by statistical error, hence we expect a substantial improvement as we will include the new generation of CLS gauge configurations  with finest values of the lattice spacing and more chiral ensembles in the following stage of the project.

	\acknowledgments
	\noindent
	We acknowledge PRACE and RES for giving us access to computational resources at MareNostrum (BSC). We thank CESGA for granting access to Finis Terrae II. This work is supported by the European Union's Horizon 2020 research and innovation programme under grant agreement No 813942 and by the Spanish MINECO through project PGC2018-094857-B-I00, the Centro de Excelencia Severo Ochoa Programme through SEV-2016-0597
	and the Ramón y Cajal Programme RYC-2012-0249. We are grateful to CLS members for producing the gauge configuration ensembles used in this study.

\end{document}